\documentclass{ieeeaccess}
\usepackage{cite}
\usepackage{amsmath,amssymb,amsfonts, bm}
\usepackage{algorithmic}
\usepackage{graphicx}
\usepackage{textcomp}
\usepackage[caption=false]{subfig}
\def\BibTeX{{\rm B\kern-.05em{\sc i\kern-.025em b}\kern-.08em
    T\kern-.1667em\lower.7ex\hbox{E}\kern-.125emX}}
\usepackage{cuted}
\usepackage[font=small,tableposition=top]{caption}
\usepackage[braket]{qcircuit}
\usepackage{url}
\usepackage[breaklinks]{hyperref}
\usepackage{ulem}

\hypersetup{
  breaklinks=true,
  colorlinks=true,
  linkcolor=blue,
  urlcolor=blue,
  citecolor=blue
}

\newcommand{\beq}[0]{\begin{equation}}
\newcommand{\eeq}[0]{\end{equation}}
\newcommand{\inlinecode}{\texttt}

\begin{document}
\history{Date of publication xxxx 00, 0000, date of current version xxxx 00, 0000.}
\doi{10.1109/TQE.2020.DOI}

\title{Benchmarking Hamiltonian Noise in the D-Wave Quantum Annealer}
\author{\uppercase{Tristan Zaborniak}\authorrefmark{1},
\uppercase{Rog\'{e}rio de Sousa\authorrefmark{1, 2}}
}
\address[1]{Department of Physics and Astronomy, University of Victoria, Victoria, British Columbia V8W 2Y2, Canada}
\address[2]{Centre for Advanced Materials and Related Technology, University of Victoria, Victoria, British Columbia V8W 2Y2, Canada}
\tfootnote{This work was supported by NSERC (Canada) through its Discovery program (Grant numbers RGPIN-2015-03938 and RGPIN-2020-04328).}

\markboth
{Zaborniak \headeretal: Benchmarking Hamiltonian noise in the D-Wave Quantum Annealer}
{Zaborniak \headeretal: Preparation of Papers for IEEE Transactions on Quantum Engineering}

\corresp{Corresponding author: Rog\'{e}rio de Sousa (email: rdesousa@uvic.ca).}

\begin{abstract}

Various sources of noise limit the performance of quantum computers by altering qubit states in an uncontrolled manner throughout computations and reducing their coherence time. In quantum annealers, this noise introduces additional fluctuations to the parameters defining the original problem Hamiltonian, such that they find the ground states of problems perturbed from those originally programmed. Here we describe a method to benchmark the amount of noise affecting the programmed Hamiltonian  of a quantum annealer. We show that a sequence of degenerate runs with the coefficients of the programmed Hamiltonian set to zero leads to an estimate of the noise spectral density affecting Hamiltonian parameters {\it ``in situ''} during the quantum annealing protocol. The method is demonstrated in D-Wave's  lower noise 2000 qubit device (\inlinecode{DW\_2000Q\_6}) and in its recently released 5000 qubit device (\inlinecode{Advantage\_system1.1}). Our benchmarking of \inlinecode{DW\_2000Q\_6} shows Hamiltonian noise dominated by the $1/f^{0.7}$ frequency dependence characteristic of flux noise \emph{intrinsic to the materials} forming flux qubits. In contrast, \inlinecode{Advantage\_system1.1} is found to be affected by additional noise sources for low annealing times, with underlying intrinsic flux noise amplitudes $2-3$ times higher than in  \inlinecode{DW\_2000Q\_6} for all annealing times.

\end{abstract}


\titlepgskip=-15pt

\maketitle

\section{Introduction}

Quantum annealers are non-universal quantum computers designed to solve optimization problems \cite{Johnson2011, Boixo2014, Albash2018}. 
Because quantum annealing (QA) has less stringent requirements on qubit control than gate-based quantum computing, it permits an easier route for scaling up to a large number of qubits. The company D-Wave Systems Inc. currently offers access to its 2000 and 5000 qubit quantum annealers on its cloud-based platform D-Wave Leap \cite{DWaveLeap}.

The D-Wave quantum annealer (hereafter annealer) is a lattice of superconducting quantum interference device (SQUID) flux qubits \cite{Harris2010}. It realizes the following programmable quantum Hamiltonian: 
\beq
\begin{aligned}
{\cal H}_{{\rm QA}}&=-\frac{A(t)}{2}\sum_i \sigma_{x}^{(i)}\\
&+ \frac{B(t)}{2}\left[\sum_i h_i\sigma_{z}^{(i)}+\sum_{i>j}J_{ij}\sigma_{z}^{(i)} \sigma_{z}^{(j)} \right],
\label{H_QA}
\end{aligned}
\eeq
where $\sigma_{x}^{(i)}$ and $\sigma_{z}^{(i)}$ are $x$ and $z$ Pauli matrices acting on qubit $i$, satisfying $\sigma_{z}^{(i)}\ket{\pm 1}=\pm \ket{\pm 1}$ (it is convenient to denote the two qubit states by $\{\ket{+1},\ket{-1}\}$, instead of the usual $\{\ket{0},\ket{1}\}$).  The parameter $h_i$ is called the ``bias" on qubit $i$, and  $J_{ij}$ is called the ``coupling" between qubits $i$ and $j$. The $h_i$ and $J_{ij}$ are dimensionless programmable input parameters that define the problem to be solved. 
Time-dependent constants $A(t)$ and $B(t)$ are the energies that determine the annealing schedule. At $t=0$, $A(0)\gg B(0)$ with $A(0)/k_B\approx 0.3$~K. Since the operating temperature is $0.012$~K \cite{DWaveTechQPU}, the system relaxes towards the ground state of the $n$-qubit system, which is in the uniform superposition of all computational basis states $\ket{\psi_0(0)}=\left[\left(\ket{+1}+\ket{-1}\right)\otimes \cdots \otimes \left(\ket{+1}+\ket{-1}\right)\right]/\sqrt{2^n}$. As time evolves, the device decreases $A(t)$ and increases $B(t)$, so that at time $t=t_a$ (the end of the anneal schedule) $A(t_a)\ll B(t_a)$, with $B(t_a)/k_B\approx 0.6$~K \cite{DWaveTechQPU}. Under circumstances approaching satisfaction of the adiabatic theorem \cite{Albash2018}, the system remains in the ground state throughout the time evolution, and each qubit state in $\ket{\psi_0(t_a)}$ can be read out to determine the lowest energy solution of the Ising Hamiltonian:
\beq
{\cal H}_{{\rm Ising}}(\bm{s})=\sum_i h_i s_i +\sum_{i>j}J_{ij}s_i s_j,
\label{H_Ising}
\eeq
where $s_i=\pm 1$ are (classical) Ising spin variables. As it turns out, the problem for finding the vector $\bm{s}$ that minimizes Eq.~(\ref{H_Ising}) belongs to the NP-Hard complexity class of classical computation, and maps with polynomial overhead to a number of important combinatorial problems including max-cut, number partitioning, exact cover, and 3-satisfiability \cite{Farhi2001, Lucas2014}. 

To date there is no consensus on which choice of parameters for Eq.~(\ref{H_Ising}) should be used to benchmark the performance of QAs with different physical hardware specifications \cite{Perdomo-Ortiz2019}. It seems that success probabilities for reaching the global energy minimum depend crucially on the hardware topology and varies wildly over the range of coefficients $h_i, J_{ij}$ \cite{Katzgraber2014}.

Similar to all architectures for quantum computing, QA hardware is sensitive to environmental noise. The theory of open quantum systems shows that noise during the annealing process leads to transitions out of the ground state \cite{Ashhab2006, Albash2012, Smirnov2018}. However, under certain circumstances, noise can also help the system return to the ground state, leading to higher success probability \cite{Childs2001, Amin2008}. 

Given that it has been difficult to directly characterize the full impact of noise on QAs and properly benchmark its accuracy for ground state estimation, we shall take an indirect approach. Here we argue that at an important part of the loss of accuracy in QAs is related to noise-induced fluctuations of the parameters $h_i, J_{ij}$ defining the problem Hamiltonian. We introduce the concept of ``Hamiltonian noise'' and show that it is possible to benchmark with a simple protocol that is generally applicable to all QA architectures.

The coefficients of the problem Hamiltonian Eq.~(\ref{H_Ising}) are set by a large number of noisy electronic devices. As a result, the bias and coupling coefficients of the problem Hamiltonian at $t$ suffer the additional contributions of the noise amplitudes, and become $h_i+\phi_i(t)$ and $J_{ij}+\theta_{ij}(t)$, where $h_i,J_{ij}$ are the originally programmed coefficients, and $\phi_i(t),\theta_{ij}(t)$ are random stochastic processes.
In light of measurements suggesting $\theta_{ij}(t)< \phi_i(t)$ (Figs. 3.1 and 3.2 in \cite{DWaveTechQPU}), our paper focuses on the impact of $\phi_i(t)$. Several mechanisms contribute to $\phi_i(t)$: cross-talk with adjacent qubits, intrinsic flux noise in the SQUIDs forming qubits, finite step size in the control hardware defining $h_i$, finite bandwidth in the analog signals during anneal, non-identical qubits with different persistent currents, temperature  fluctuations, and high-energy photon flux \cite{DWaveTechQPU}. 

A notable consequence of $\phi_i(t)\neq 0$ is that the system ends up solving the ``wrong problem''  even when the QA finds the correct ground state \cite{Pearson2019}. We refer to this effect as ``Hamiltonian noise''.

D-Wave provides quantitative information on Hamiltonian noise (which they call ``Integrated Control Errors'') in Chapter 3 of its technical description of the QPU \cite{DWaveTechQPU}. The company uses a series of measurements on two Ising spins $s_i, s_j$ to determine a phase diagram for their ensemble averages as a function of a range of input parameters $h_i,h_j,J_{ij}$. Comparing this phase diagram to thermal averages of the ideal Hamiltonian Eq.~(\ref{H_Ising}) allows them to estimate the shifts $\phi_i$, $\phi_j$ and $\theta_{ij}$ in each pair of qubits. Doing this for every coupled $i,j$ pair in the architecture yields the ensemble averages over all qubits $\langle \phi\rangle$, $\langle \theta\rangle$ and the corresponding rms deviations $\sqrt{\langle \phi^2\rangle}$ and $\sqrt{\langle \theta^2\rangle}$.

Here we propose an alternative method to benchmark Hamiltonian noise using the readout outcomes of D-Wave annealers. The method is based on our recent report describing {\it in situ} characterization of the frequency dependence of the noise affecting QAs \cite{Zaborniak2020}. While the procedure for randomized benchmarking of quantum gate fidelity is well established in gate-model quantum computers \cite{Magesan2011}, we are not aware of any protocols for the user to benchmark the stability of the parameters defining the qubit Hamiltonian, neither for gate model nor for quantum annealing architectures. The method we propose is generally applicable to all quantum annealing architectures. The method makes evident the benefits of benchmarking noise across a wider frequency range, rather than the usual approach of focusing on static (zero-frequency) averages. As we shall show, this feature enables diagnosis of the noise source if its frequency dependence comes out similar to well-characterized physical mechanisms.

\begin{figure*}
\centerline{\includegraphics[width=0.99\textwidth]{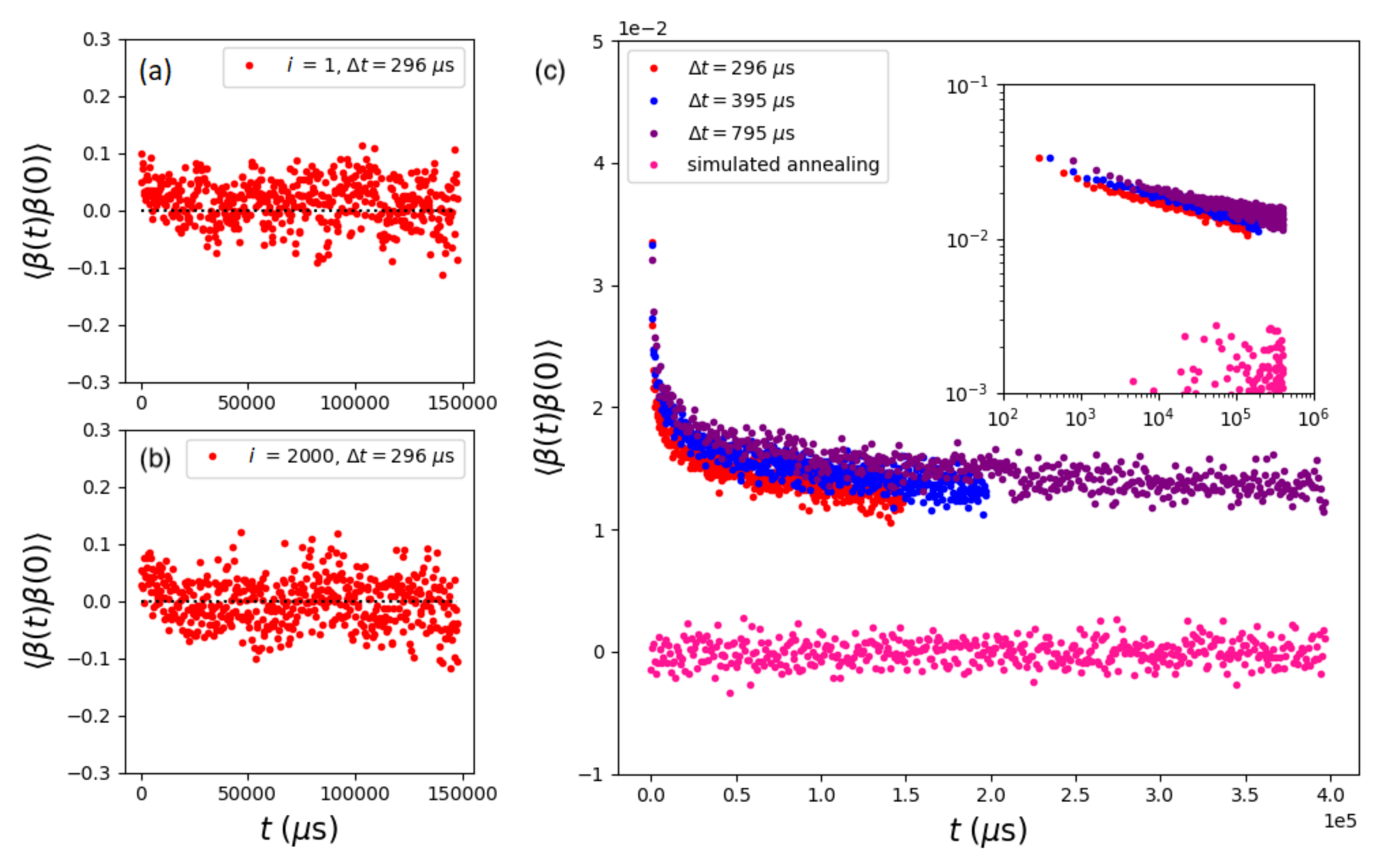}}
\caption{Solution-state time correlation function obtained from $N=1000$ degenerate QA runs in the \inlinecode{DW\_2000Q\_6} QPU. (a) Correlation function for qubit $i=1$, for annealing time $t_a=1$~$\mu$s ($\Delta t = 296$ $\mu$s). (b) Same for qubit $i=2000$. (c) Correlation functions after averaging over all 2000 qubits in lin-lin and log-log (inset), for $t_a=1, 100, 500$~$\mu$s. For comparison we also show the correlation function obtained by sampling the same degenerate problem by classical simulated annealing using \inlinecode{neal} \cite{DWaveLeap}.}
\label{betacorr}
\end{figure*}

\section{Degenerate Quantum Annealing as a witness for Hamiltonian noise}

We now describe a series of annealing experiments using the D-Wave Leap interface with the \inlinecode{DW\_2000Q\_6} quantum processing unit (QPU). Later, we apply the same method to \inlinecode{Advantage\_system1.1}.
Our method is to run $N$ instances of the \emph{degenerate} Ising problem ($h_i=J_{ij}=0$), and each time record the readout outcome of the  $n$ qubits. Each run yields a set of $n$ time series $\beta_i(j)=\pm 1$, denoting the  state of qubit $i=1,\ldots, n$ readout at time $t_j=(j+1)\Delta t$, where $j=0,\ldots, N-1$. Here $\Delta t = t_{d} + t_a$, where $t_d$ is the time it takes for initialization and readout, and $t_a$ is the annealing time set by the user ($t_{d}=295$~$\mu$s and $149$~$\mu$s for \inlinecode{DW\_2000Q\_6} and \inlinecode{Advantage\_system1.1}, respectively). 

The time correlation function for the solution-state of qubit $i$ at time $t_k=k\Delta t$ is computed from:
\beq
\langle \beta(t_k)\beta(0)\rangle_i = \frac{1}{N-k}\sum_{j=0}^{N-k-1}\beta_i(j+k)\beta_i(j).
\eeq
Clear from this definition is that as $k$ increases, fewer observations of $\beta_i(j+k)\beta_i(j)$ are used in calculating $\langle\beta(t_k)\beta(0)\rangle_i$, with only one observation being used in the case that $k = N-1$. In our calculations, we chose to consider $k = 0, 1, 2, ..., N/2$ such that the maximum and minimum numbers of observations used in computing $\langle\beta(t_k)\beta(0)\rangle_i$ differ by only a factor of 2. With $N = 1000$, this means that we use a minimum of 500 observations in computing $\langle\beta(t_k)\beta(0)\rangle_i$. Figures~\ref{betacorr}(a)~and~\ref{betacorr}(b) shows the measured solution-state time correlation function for qubits $i=1$ and $i=2000$ of \inlinecode{DW\_2000Q\_6}, both for annealing time $t_a=1$~$\mu$s ($\Delta t=296$~$\mu$s). 

A significantly smoother version of the time correlation function is obtained by averaging over all $n=2000$ qubits:
\begin{equation}
    \langle \beta(t_k)\beta(0)\rangle = \frac{1}{n}\sum_{i=1}^{n}\langle \beta(t_k)\beta(0)\rangle_i.
\end{equation}
Figure~\ref{betacorr}(c) shows this result for three different annealing times, $t_a=1, 100, 500$~$\mu$s, corresponding to $\Delta t=296, 395, 795$~$\mu$s. For comparison to what would be true in the absence of Hamiltonian noise, we also include the correlation function obtained by running the same degenerate problem 
by way of classical simulated annealing using \inlinecode{neal} (Fig.~\ref{betacorr}(c)).  Clearly, Fig.~\ref{betacorr} reveals the presence of a long time tail in QA measurements. This is clear evidence for the presence of \emph{colored noise} in the QPU, in contrast to the uncorrrelated white noise of the simulator. 

\begin{figure}
\centerline{\includegraphics[width=9 cm]{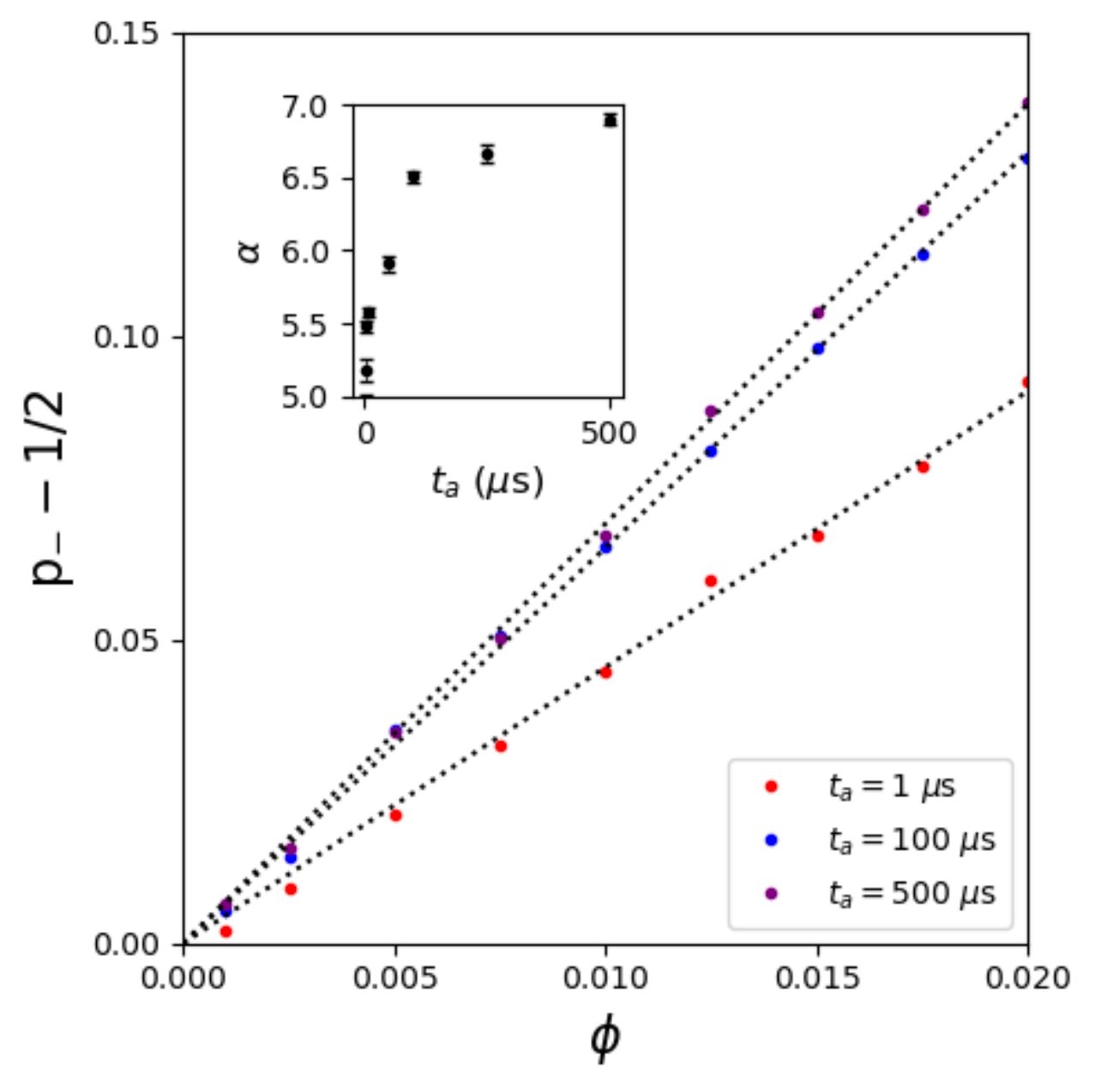}}
\caption{Probability $p_{-}$ for obtaining the one-qubit ground state $\ket{-}$ for small values of $h_i=\phi>0$ in the \inlinecode{DW\_2000Q\_6} QPU. The linear relation $p_{\pm}=\frac{1}{2}\mp \alpha(t_a)\phi$ holds with coefficient $\alpha(t_a)$ depending on the annealing time $t_a$ (inset).}
\label{ppsi0}
\end{figure}

Our interpretation of this result is that the actual $h_i$ parameters in the QPU differ from zero at readout.
They are instead given by $h_i=\phi_i(t)$, where $\phi_i(t)$ is a stochastic process. In order to characterize the details of this process, we need to find the relationship between $h_i$ and the probability for measuring the corresponding one-qubit ground state in the QA (e.g. $p_{-}=p(\ket{-1})$ for $h_i>0$) as a function of anneal time. To find $p_{\pm}$ experimentally, we performed $N=1000$ ``non-degenerate'' QA runs for several small values of $h_i=\phi >0$. We estimated $p_{\pm}$ as the fraction of runs that the annealer measured $+1$ or $-1$ for the qubit state. The result is shown in Fig.~\ref{ppsi0} where we see that the linear relation 
\beq
p_{\pm}=\frac{1}{2}\mp \alpha(t_a)\phi
\label{p_pm}
\eeq
holds for $\phi$ smaller than $10^{-2}$ \cite{NoteEq5}. Values for the coefficient $\alpha(t_a)$ depend strongly on the annealing time $t_a$, as shown in the inset.

\begin{figure}
\centerline{\includegraphics[width=9cm]{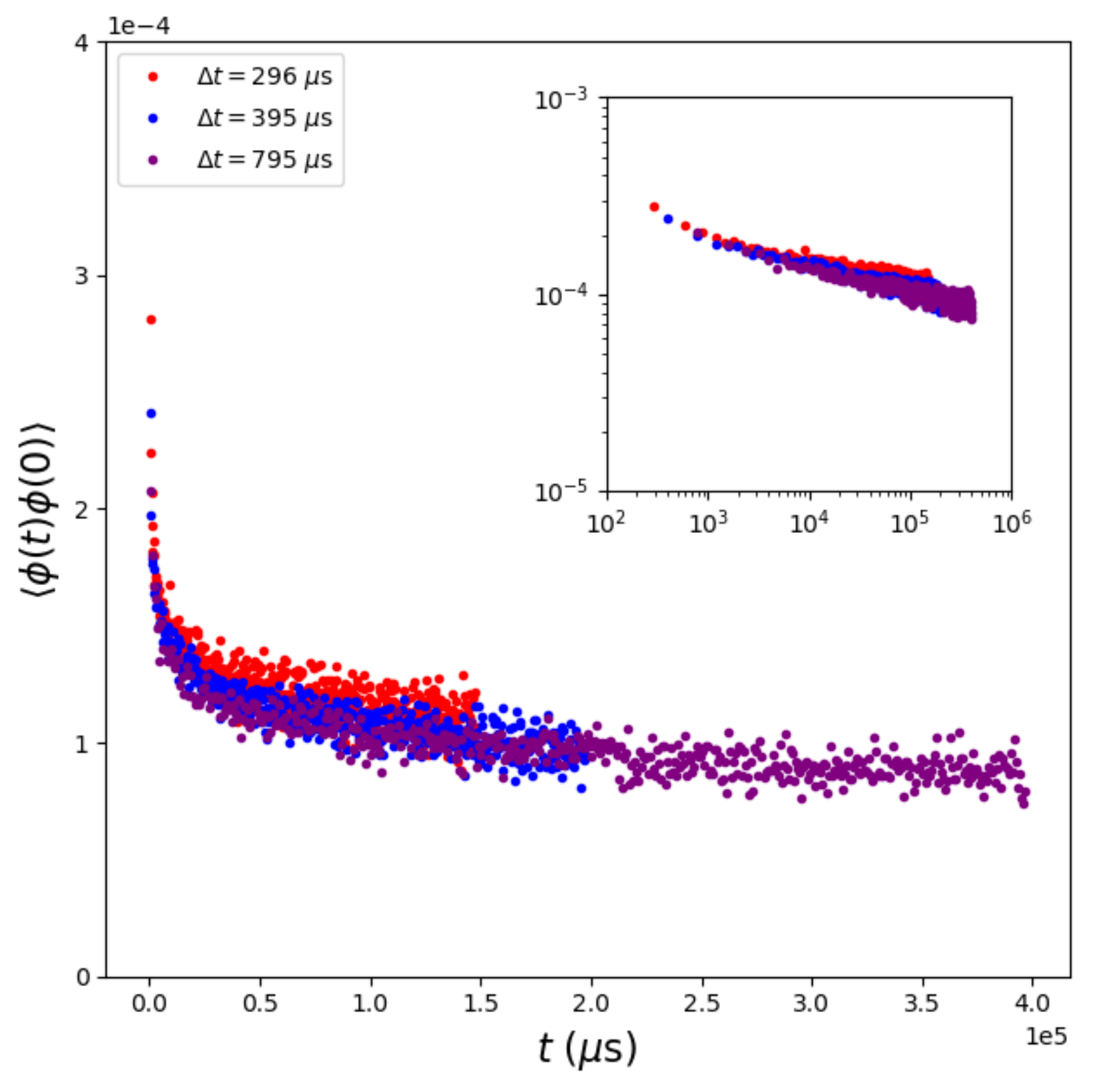}}
\caption{Time correlation function for $h_i=\phi$  in the \inlinecode{DW\_2000Q\_6} QPU, averaged over all qubits, obtained from $\langle\phi(t)\phi(0)\rangle=\langle \beta(t)\beta(0)\rangle /[4\alpha^{2}(t_a)]$, see Eq.~(\ref{beta_phi_relation}). Data shown for $t_a=1,100,500$~$\mu$s. The log-log plot is shown in the inset.}
\label{phicorr}
\end{figure}

\begin{figure*}
\begin{center}
\subfloat[Noise spectral density in the \inlinecode{DW\_2000Q\_6}.]{\includegraphics[width=0.49\textwidth]{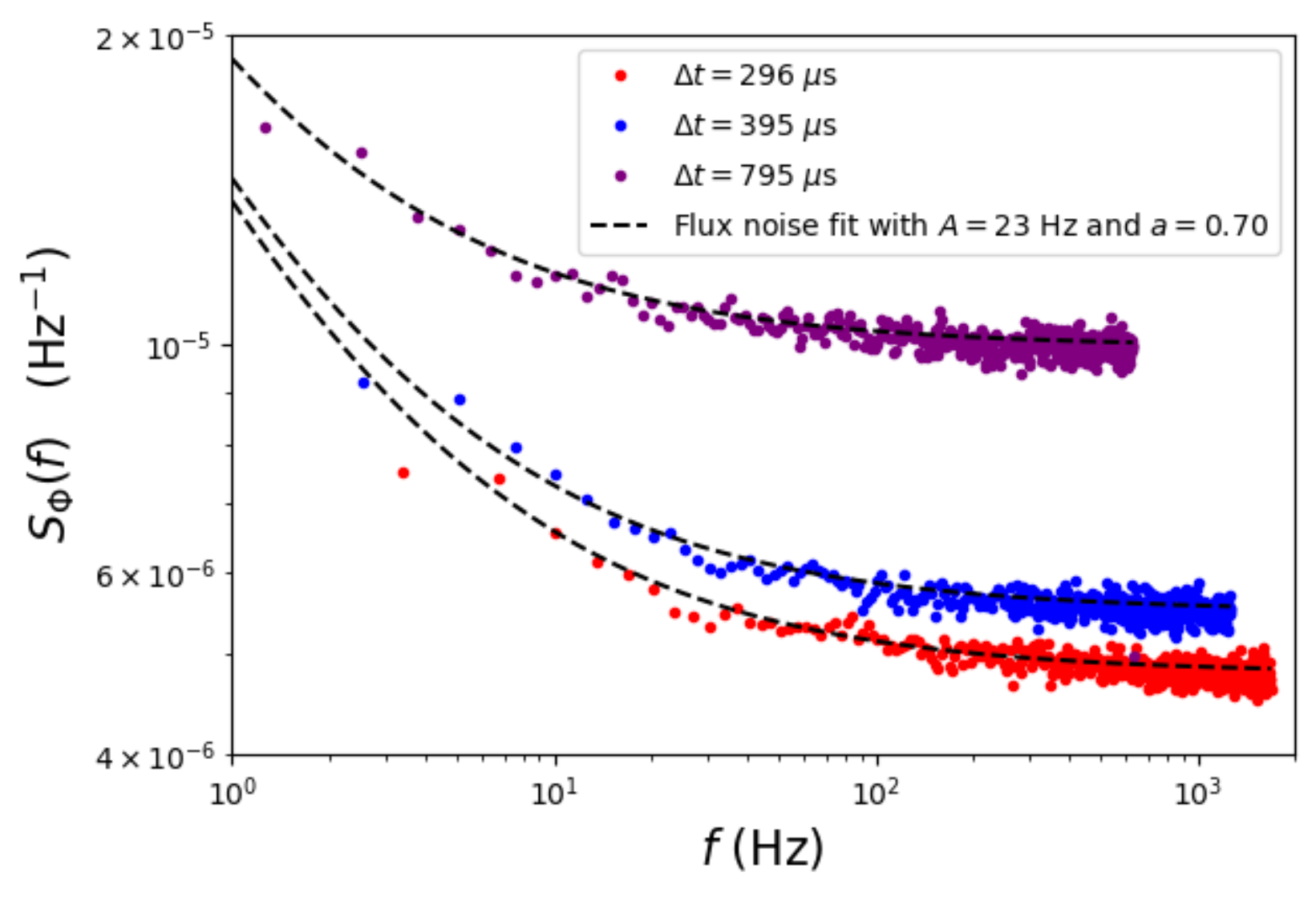}
\label{fignoise}}
\subfloat[Noise spectral density in the \inlinecode{Advantage\_system1.1}.]{\includegraphics[width=0.49\textwidth]{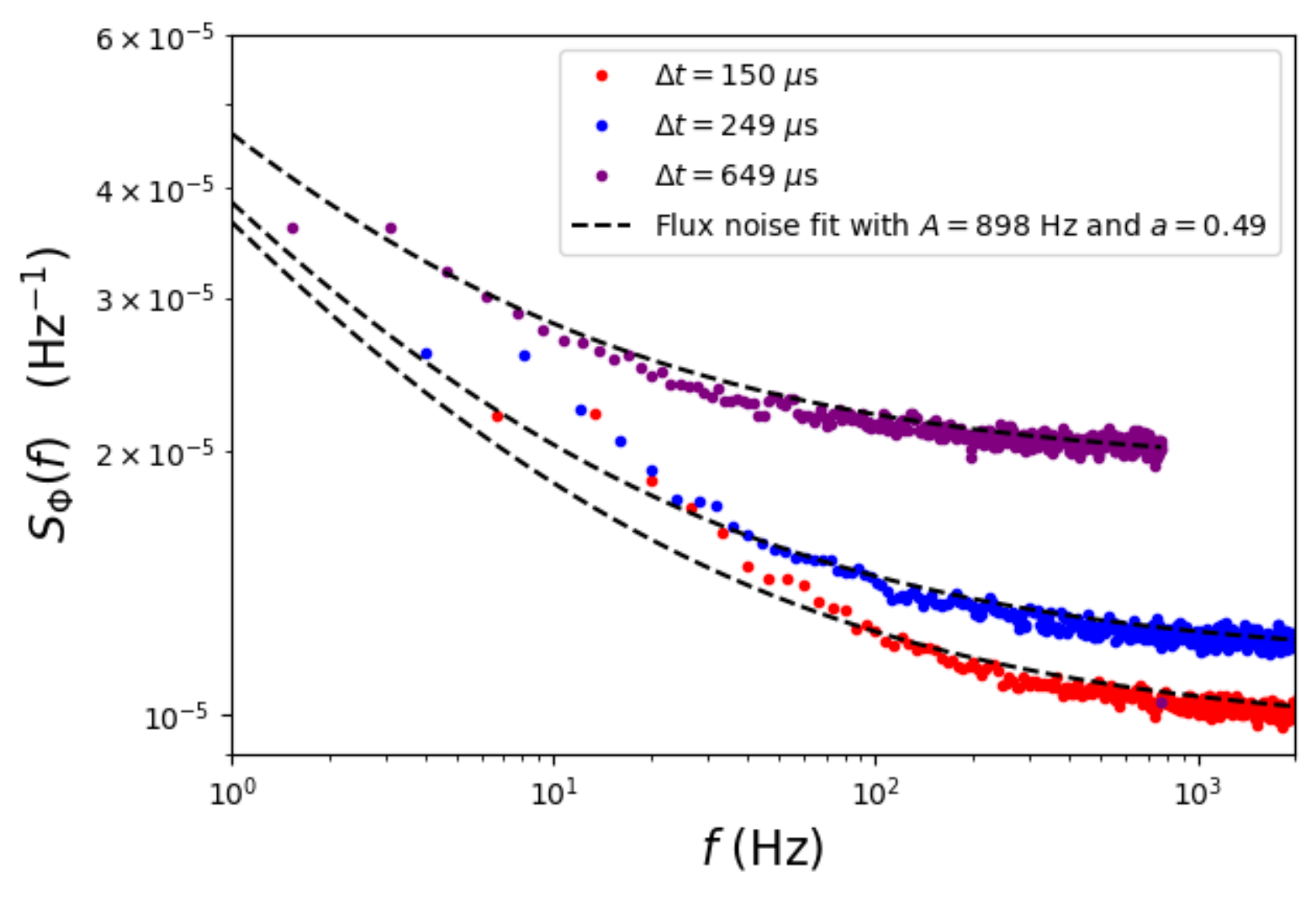}
\label{fignoiseAdvantage}}
\end{center}
\caption{Estimated noise spectral density in the D-Wave systems using the Welch method. The dashed lines are global fits based on Eq.~(\ref{model}) using two fitting parameters independent of $\Delta t$: the flux noise amplitude $A$ and exponent $a$ in $\left(A/f\right)^{a}$, plus the artifactual white noise background given by Eq.~(\ref{W_eqn}).
(a) \inlinecode{DW\_2000Q\_6}: The flux noise fits are quite good for all annealing times, showing that intrinsic flux noise dominates over all other sources of noise.
(b) \inlinecode{Advantage\_system1.1}: Here the flux noise fit is only good for the longest annealing time, $\Delta t=649$~$\mu$s. The shorter annealing times $\Delta t=150, 249$~$\mu$s clearly deviate from the intrinsic flux noise law $\left(\frac{A}{f}\right)^{a}$ in the low frequency range. This indicates that additional noise mechanisms are playing an important role.}
\end{figure*}

Armed with Eq.~(\ref{p_pm}), we get an explicit relation between the correlation functions for $\beta(t)$ and $\phi(t)$:
\begin{align}
\langle \beta(t)\beta(0)\rangle  
&=\left\langle\left[p_+(t)-p_-(t)\right]\left[p_+(0)-p_-(0)\right]\right\rangle\nonumber\\
&= 4\alpha^2(t_a)\langle\phi(t)\phi(0)\rangle.
\label{beta_phi_relation}
\end{align}
We should remark that this relation is based on the assumption that QA was used to probe $\phi$; as a result, Eq.~(\ref{beta_phi_relation}) requires $t\ge\Delta t$. It does not hold for $t=0$, where $\langle\beta(t)\beta(0)\rangle=1$ by design, such that $\langle\phi^2\rangle$ is not necessarily equal to $1/[4\alpha^2(t_a)]$. Furthermore, we note that under the assumption that $\phi(t)$ is Gaussian-distributed, all higher order correlation functions of its time series can be written as products of $\langle\phi(t)\phi(0)\rangle$.

Figure~\ref{phicorr} shows the resulting $\langle \phi(t)\phi(0)\rangle$ obtained using Eq.~(\ref{beta_phi_relation}) with $\alpha(t_a)$ adjusted within their ranges of uncertainty to yield a better collapse of the three different $t_a$ curves. 

\section{Estimated power spectral density}

We now obtain an estimate for the symmetrized noise spectral density associated with $\phi$: 
\beq
S_\phi(f)=\int_{-\infty}^{\infty} dt \textrm{e}^{2\pi i f t} \left[\langle \phi(t)\phi(0)\rangle + \langle \phi(0)\phi(t)\rangle\right].
\label{S_phi}
\eeq
To do this, we applied the Welch method \cite{Welch1967} directly to the $\beta_i(j)$  time series, averaged over all qubits $i$, and divided by $4\alpha^2(t_a)$ following Eq.~(\ref{beta_phi_relation}).  The output of the Welch method is a smoothed estimate of the periodogram:
\begin{equation}
P(f_l)=\frac{2\Delta t}{N}\left\langle \left|\tilde{\phi}(f_l)\right|^{2}\right\rangle,
\end{equation}
where $\tilde{\phi}(f_l)=\sum_{j=0}^{N-1}\textrm{e}^{-2\pi i f_l t_j}\phi(t_j)$, for frequencies $f_l=l/(N\Delta t)$ with $l=0,1,\ldots, N/2$. 
We used the largest possible segment length for smoothing (\inlinecode{nperseg} $=N=1000$, the length of the time series \cite{WelchPython}) to minimize the usual periodogram oscillations in the low frequency region. The result  is shown in Fig.~\ref{fignoise}. 

As pointed out above, the fact that $\langle \beta^2\rangle=1$ gives an additional contribution to the correlation function $\langle\phi(t)\phi(0)\rangle$  at $t=0$. Because this contribution is present only at $t=0$, it contributes an artifactual white noise background to $S_\phi(f)$. 
This white noise background enforces the satisfaction of the sum rule:
\beq
\int_{0}^{\frac{1}{2\Delta t}}df S_{\phi}(f)={\rm lim}_{t\rightarrow 0}\langle\phi(t)\phi(0)\rangle = \frac{1}{4\alpha^{2}(t_a)},
\label{sumrule}
\eeq
where we used the fact that the total noise power from the Welch periodogram is contained in the $[0, 1/(2\Delta t)]$ frequency interval. 

With Eq.~(\ref{sumrule}) we are able to obtain a simple model that explains our measurements. We assume the noise is \emph{dominated} by \emph{intrinsic magnetic flux noise} within the materials forming the SQUIDs \cite{Wellstood1987, Lanting2014a, Laforest2015, Quintana2017} plus the artifactual white noise background:
\beq
S_\phi(f)= \left[\left(\frac{A}{f}\right)^{a} + W\right]\mu{\rm s},
\label{model}
\eeq
where the amplitude $A$ and exponent $a$ are fitting parameters independent of $t_a$ and $\Delta t$. The white noise background $W$ is calculated using Eq.~(\ref{sumrule}) to be:
\beq
W=\frac{(\Delta t)}{2\alpha^2(t_a)\mu{\rm s}}-\frac{\left[2(\Delta t)A\right]^{a}}{1-a}.
\label{W_eqn}
\eeq
In order to fit the data we first removed the $f=0$ data points from the power spectral densities estimated with the Welsh method, so that any static effect such as long-lived spin polarization \cite{Lanting2020}  is subtracted from the data (note that Eq.~(\ref{S_phi}) did not subtract $\langle \phi(t)\rangle$ from the fluctuations of $\phi$). Equations~(\ref{model})~and~(\ref{W_eqn}) were used to fit all three curves simultaneously in Fig.~\ref{fignoise} \cite{FitMethod} leading to $A=(23\pm 1)$~Hz and $a=(0.70\pm 0.02)$ (dashed lines) for the  \inlinecode{DW\_2000Q\_6}.

Figure~\ref{fignoiseAdvantage} shows the same analysis applied to the newer 5000 qubit device, \inlinecode{Advantage\_system1.1}. Here we see a clear deviation from the intrinsic flux noise law $\left(A/f\right)^{a}$ in the low frequency range for the lowest annealing times $\Delta t=150, 249$~$\mu$s. This provides evidence that other sources of noise (other than intrinsic flux noise) are playing a role in  \inlinecode{Advantage\_system1.1}. There are many possibilities: Much higher $J_{ij}$ noise (due to the larger connectivity), cross talk, photon loss, etc. In contrast, data taken with the longest annealing time $\Delta t=649$~$\mu$s seems to follow the intrinsic flux noise expression in the whole frequency range, suggesting that once again intrinsic flux noise dominates. However, it does so with larger amplitude: $S_{\phi}(10~{\rm}$ Hz) for  \inlinecode{Advantage\_system1.1} is $2-3$ times larger than for \inlinecode{DW\_2000Q\_6}. A simultaneous fit to all three curves yields  $A=(900\pm 80)$~Hz with a lower exponent $a=(0.49\pm 0.01)$. 

\section{Conclusions}

In conclusion, we describe a method for benchmarking Hamiltonian noise in QA hardware, based on sampling the states of qubits after annealing the device in the degenerate regime. 
Applying this method to D-Wave's \inlinecode{DW\_2000Q\_6} and  \inlinecode{Advantage\_system1.1} yields a comparison for the amount of noise in two different device architectures, based on estimates for the qubit bias time correlation function $\langle \phi(t)\phi(0)\rangle$ and its noise spectral density $S_\phi(f)$. 

Our measurements of the time correlation function for the qubit bias noise $\phi(t)$ lead to an estimate for the uncertainty when programming bias $h_i$ in QA. The rms uncertainty in \inlinecode{DW\_2000Q\_6}
is found to be
$\sqrt{\langle \phi^2\rangle}\approx \sqrt{\langle \phi(\Delta t)\phi(0)\rangle}= 2\times 10^{-2}$ when the smallest $\Delta t = 296$~$\mu$s is used.   
This is comparable to the estimate of $\textrm{ Max}\{\sqrt{\langle \phi^2\rangle}\}= 2.4\times 10^{-2}$
obtained by D-Wave from fitting thermal averages to the QPU outputs over a range of nonzero $h_i$ and $J_{ij}$ (See Fig.~3.1 of \cite{DWaveTechQPU}). 

Our measurements for \inlinecode{Advantage\_system1.1} yield $\sqrt{\langle \phi^2\rangle}\approx \sqrt{\langle \phi(\Delta t)\phi(0)\rangle}= 4\times 10^{-2}$ for the smallest $\Delta t = 150$~$\mu$s, a value that is two times larger than in \inlinecode{DW\_2000Q\_6}. Given that estimates for $\textrm{ Max}\{\sqrt{\langle \phi^2\rangle}\}$ over a range of $h_i, J_{ij}$ are not available, it would be prudent to investigate whether our estimate holds away from the degenerate regime. We should emphasize that it is  straightforward to generalize our benchmark protocol for $h_i \neq 0$ and $J_{ij}\neq 0$. Doing so would allow one to rigorously check whether or not the affecting noise level is the same away from the $h_i=J_{ij}=0$ point.

Moreover, the estimation of the noise spectral density $S_{\phi}(f)$ for different annealing times $t_a$ allows us to conclude that in \inlinecode{DW\_2000Q\_6} the qubit bias noise is dominated by the $1/f^{0.7}$ frequency dependence characteristic of flux noise \emph{intrinsic to the materials} forming the device  \cite{Wellstood1987, Lanting2014a, Quintana2017}. This shows that in \inlinecode{DW\_2000Q\_6} intrinsic flux noise currently plays a more important role than other mechanisms such as cross-talk due to adjacent qubits and control hardware.

\inlinecode{Advantage\_system1.1} is found to be affected by underlying intrinsic flux noise of amplitude $2-3$ times higher than \inlinecode{DW\_2000Q\_6} for all annealing times, with an additional low-frequency noise contribution for low annealing times.
The additional noise sources may be related to the greater connectivity of the device, either due to additional fluctuations in $J_{ij}$ or cross-talk. 

\section*{Acknowledgment}
We thank M. Amin, T. Lanting, A. Nava, and S. Tomkins for useful discussions and critical reading of the manuscript. This work was supported by NSERC (Canada) through its Discovery program (Grant numbers RGPIN-2015-03938 and RGPIN-2020-04328).

\bibliographystyle{IEEEtranDOI}
\bibliography{IEEEabrv, HnoiseTQE}

\begin{thebibliography}{10}
\bibitem{NoteEq5}
We remark that the linear $\phi$ dependence in Eq.~(\ref{p_pm}) plays a crucial
  role in our method. The linearity also holds when $h_i\neq 0$, albeit with
  different coefficients $\alpha(t_a, h_i)$. For more complicated problem
  Hamiltonians $p_{\pm}$ may show differing coefficients for $\phi>0$ and
  $\phi<0$. For these cases Eq.~(\ref{beta_phi_relation}) will not hold, and
  our method will have to be modified.
\end{thebibliography}

\EOD

\end{document}